\documentclass[a4paper,11pt]{article}
\usepackage{jinstpub} 
\usepackage{lineno}
\usepackage{threeparttable}
\usepackage{cancel}
\usepackage{xcolor}

\title{\boldmath The Intrinsic Energy Resolution of LaBr$_{3}$(Ce) Crystal}

\author[a,b,*]{Pei-Yi Feng,\note[*]{Corresponding author.}}
\author[c,*]{Xi-Lei Sun,}
\author[a]{Zheng-Hua An,}
\author[d]{Cheng-Er Wang,}
\author[a]{Da-Li Zhang,}
\author[a]{Xin-Qiao Li,}
\author[a,b]{Chao Zheng,}
\author[a]{Shao-Lin Xiong,}
\author[a]{Hong Lu}

\affiliation[a]{Key Laboratory of Particle Astrophysics,\\ Institute of High Energy Physics, Chinese Academy of Sciences,\\ Beijing 100049, China}
\affiliation[b]{University of Chinese Academy of Sciences, Chinese Academy of Sciences,\\ Beijing 100049, China} 
\affiliation[c]{State Key Laboratory of Particle Detection and Electronics,\\ Institute of High Energy Physics, Chinese Academy of Sciences,\\ Beijing 100049, China}
\affiliation[d]{National Engineering Research Center for Rare Earth, Grirem Advanced Materials Co., Ltd.,\\ Beijing 100088, China}

\emailAdd{fengpeiyi@ihep.ac.cn (PYF), sunxl@ihep.ac.cn (XLS).}

\abstract{This study aims to provide an accurate estimation of the intrinsic resolution of LaBr$_3$(Ce) crystal through a combination of experimental and simulation methods. We re-analyzed the data from previous Wide-Angle Compton Coincidence (WACC) and Hard X-ray Calibration Facility (HXCF) experiments, conducted PMT Single-Photoelectron Calibration (SPEC) and radial non-uniformity (also called Spot Scanning, SS) experiments to acquire new data, and combined these results with Geant4 simulations to isolate the contribution of each physical process to the total energy resolution, thereby allowing for a precise estimation of the scintillator's intrinsic resolution. For 100 keV X-rays, the total energy resolution of LaBr$_3$(Ce) crystal is 3.99\% $\pm$ 0.04\% (expressed as 1-$\sigma$), with statistical fluctuations and intrinsic resolution as the main components, contributing 2.47\% $\pm$ 0.00\% and 3.06\% $\pm$ 0.06\%, respectively. We identify two main sources of intrinsic resolution: one primarily due to non-proportional scintillation, contributing 2.28\% $\pm$ 0.00\%, and the other due to fluctuations in the energy transfer process, contributing 2.04\% $\pm$ 0.08\%. We quantified six components of the total energy resolution and reconstructed the photon response using Geant4. The consistency between the reconstructed relative light yield and the experimental measurements validated the mass model of the LaBr$_3$(Ce) detector used in the simulations.}



\begin{document}
\maketitle
\flushbottom


\section{Introduction}

LaBr$_3$(Ce) crystals represent a promising advancement in inorganic scintillators, boasting high light output, exceptional energy and time resolution, excellent energy linearity, and rapid luminescence decay. These characteristics of LaBr$_3$(Ce) crystals find wide applications in various fields, including studies of scintillation mechanisms \cite{canning2011first}, nuclear spectroscopy \cite{kumar2009efficiency, mazumdar2013studying, dhibar2018characterization}, nuclear medicine \cite{nassalski2006road}, three-dimensional imaging in gamma-ray astronomy \cite{gostojic2016characterization, dong2023development, 2023WW}, security and remote sensing \cite{kozyrev2016comparative}, gamma-ray logging \cite{wan2017research}, and environmental monitoring \cite{chung1988environmental}.

Energy resolution is a crucial parameter characterizing scintillator performance \cite{wen2021compact}. Intrinsic resolution is an important component of the total energy resolution \cite{4545171}. The non-proportional luminescence of the crystal is the main cause of intrinsic resolution \cite{5076032}. These conclusions were confirmed in the experiments conducted by M. Moszynski et al \cite{moszynski2002intrinsic, moszynski2004intrinsic}.
M.S. Alekhin et al. demonstrated the energy response non-proportionality of the LaBr$_3$(Ce) crystal to X-rays \cite{alekhin2013improvement}. So, what is the contribution of the non-proportionality component to the intrinsic resolution? Besides non-proportionality, what is the proportion of other sources contributing to the intrinsic resolution? Will different types of scintillators lead to different conclusions? These questions have not been addressed in previous studies.

In fact, the three effects leading to the non-proportional response are: (1) secondary X-rays and Auger electrons generated after photoelectric absorption; (2) multiple Compton scatterings resulting in the full energy absorption of gamma rays; and (3) the production of $\delta$-rays during the scattering of secondary electrons. These effects correspond to the mechanisms of interaction between the incident photons and the material. W. W. Moses et al. confirmed that these effects, coupled with a non-proportional response, degrade the energy resolution of the scintillator \cite{4545170}.


Y. Deng et al. elucidated that the intrinsic resolution of liquid scintillator for 976 keV electrons is 1.83\%. They indicated that non-proportionality in liquid scintillator has a minimal contribution to the intrinsic resolution for electrons, suggesting that fluctuations during energy transfer processes might largely account for the intrinsic resolution \cite{deng2022exploring}. The intrinsic resolution is entirely determined by the properties of the material itself. Besides non-proportional luminescence, the process of energy transfer within the scintillator to generate photons at luminescent centers is also considered as an important source of intrinsic resolution, due to the occurrence of fluctuations in this process.

The previous study shows insufficient depth in understanding the concept of intrinsic resolution. There is a scarcity of experimental measurements concerning the crystal intrinsic resolution, and the underlying physical mechanisms require further investigation. K. Sriwongsa et al. compared and discussed the intrinsic and statistical resolutions of LaBr$_3$ and LuYAP scintillators in the energy range of 356–1332 keV \cite{sriwongsa2017comparative}. However, like other published studies, these investigations merely isolated the statistical resolution from the total energy resolution, treating the remaining part as intrinsic resolution \cite{2017Intrinsic, kaintura2021energy, sriwongsa2017comparative}. This simplified approach leads to an overestimation of intrinsic resolution.

To better comprehend the limitations and potential of these scintillators, investigating the entire physical process influencing the total energy resolution remains of significant importance. In this paper, building upon existing research on the non-proportionality of LaBr$_3$(Ce) crystal in the low-energy range \cite{FPY}, we quantified the contributions of six factors—fluctuations in energy transfer, non-proportional luminescence, uneven light collection, statistical fluctuations in photoelectrons, single-photoelectron resolution, and electronic noise—towards energy resolution through a combination of experiments and Geant4 simulations in 5.1–106.6 keV energy range. The relationship between these six components and the total energy resolution is shown in Equation~\ref{eq:resolution}. We aim to clarify the concept of intrinsic resolution, analyze the relationship between energy transfer processes, non-proportional luminescence, and intrinsic resolution, and subsequently explore the sources of intrinsic resolution.

\begin{equation}\label{eq:resolution}
(\sigma/E)^2 = \delta_{trans}^2 + \delta_{non}^2 + \delta_{un}^2 + \delta_{st}^2 + \delta_{spe}^2 + \delta_{noise}^2 .
\end{equation}

In Equation~\ref{eq:resolution}, $\sigma⁄E$ represents the total energy resolution, expressed as 1-$\sigma$. $\delta_{trans}$ indicates fluctuations in the energy transfer process, $\delta_{non}$ denotes the contribution of the non-proportional luminescence, $\delta_{un}$ accounts for uneven light collection, $\delta_{st}$ represents statistical fluctuations during photon-to-photoelectron conversion at the photocathode of photomultiplier tube (PMT), $\delta_{spe}$ stands for the effect of the PMT single-photoelectron resolution on the crystal's energy resolution, and $\delta_{noise}$ signifies the contribution of dark noise to the total energy resolution. In this study, we utilized high-performance PMTs, hence the contribution of dark noise was negligibly. As $\delta_{trans}$ and $\delta_{non}$ are entirely determined by the material's intrinsic properties, we collectively term these two components as the intrinsic resolution of the crystal (Equation~\ref{eq:intrinsic resolution}).

\begin{equation}\label{eq:intrinsic resolution}
\delta_{int}^2 = \delta_{trans}^2 + \delta_{non}^2 .
\end{equation}

This study involves four experiments: Wide-Angle Compton Coincidence (WACC), Hard X-ray Ground Calibration Facility (HXCF), PMT Single-Photoelectron Calibration (SPEC), and radial non-uniformity (also called Spot Scanning, SS) experiments. The WACC and HXCF experiments were conducted to test the non-linearity of the crystals, and their data were previously collected \cite{FPY}. In this study, we re-analyzed the WACC and HXCF data, conducted the SPEC and SS experiments to acquire new data, and combined these results with Geant4 simulations to isolate the contribution of each physical process to the energy resolution, thus enabling an accurate estimation of the scintillator's intrinsic resolution.

\section{Experimental Setups}

This section introduces the experimental setups utilized in this study. The experimental setup for non-proportional light output is employed to analyze the contribution of luminescence non-proportionality to the energy resolution \cite{FPY}. The single-photoelectron calibration of PMT is employed to analyze the contribution of statistical fluctuations in photoelectrons to the energy resolution. The setup for radial non-uniformity testing is used to analyze the contribution of uneven light collection to the energy resolution.

\subsection{Experimental Setup for Non-proportional Light Yield}\label{chap:Experimental Setup for Non-proportional Light Yield}

To measure the non-proportionality of the crystal's light yield, we conducted tests and comparative studies using Compton electrons and monochromatic X-rays for the LaBr$_3$(Ce) crystal. We utilized the Wide-Angle Compton Coincidence (WACC) technique to obtain Compton electrons in the energy range of 5.1–106.6 keV and the Hard X-ray Calibration Facility (HXCF) to generate monochromatic X-ray beams in the energy range of 8–100 keV.

\begin{figure}[!htb]
\centering
\includegraphics
  [width=0.5\hsize]
  {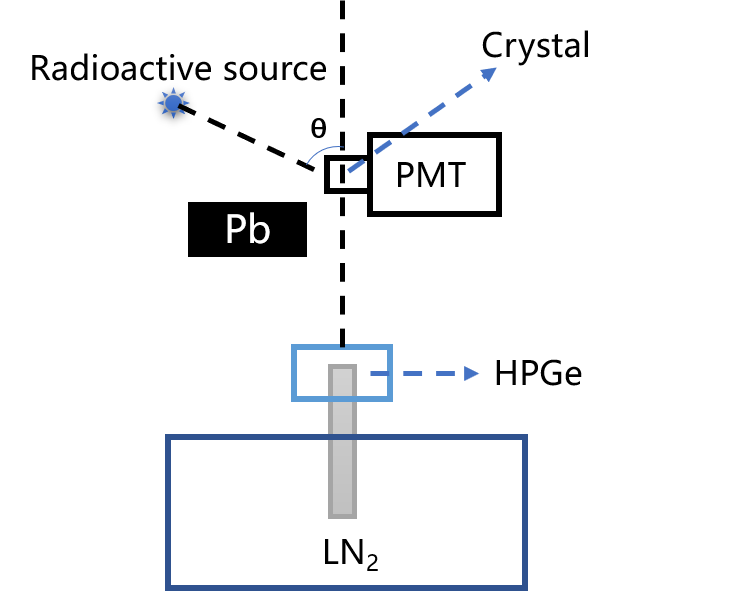}
\caption{Schematic diagram of WACC experimental setup for obtaining Compton electrons with a wide energy range \cite{FPY}.}
\label{fig:1}
\end{figure}

\begin{figure*}[!htb]
\centering
\includegraphics
  [width=\hsize]
  {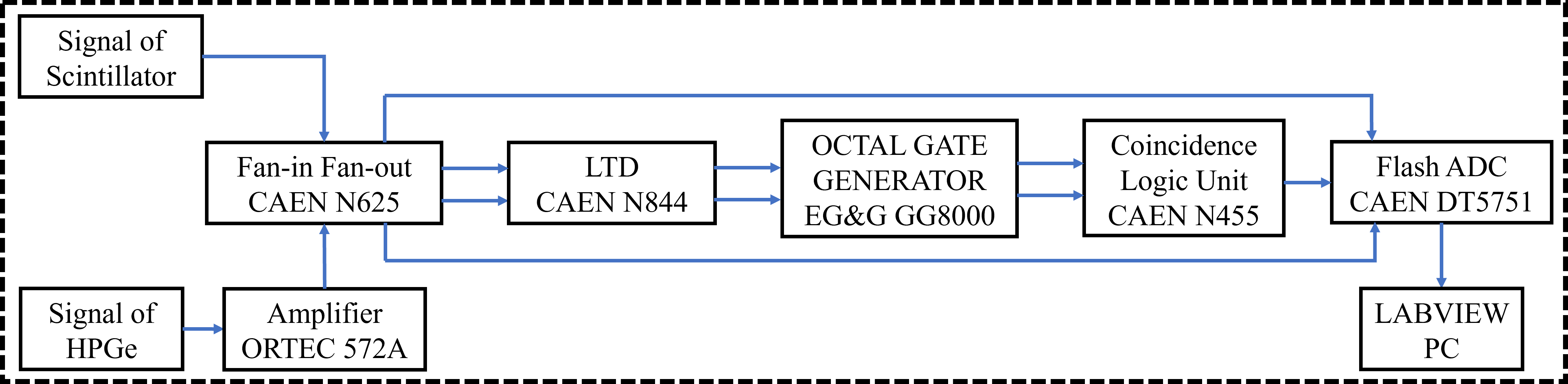}
\caption{Diagram of the data acquisition system for WACC events \cite{FPY}.}
\label{fig:2}
\end{figure*}

Figure~\ref{fig:1} depicts the WACC experimental setup, consisting of a LaBr$_3$(Ce) crystal, a high-purity Germanium (HPGe) detector, and a $^{137}$Cs radioactive source \cite{FPY}. In this study, a cylindrical LaBr$_3$(Ce) crystal with a diameter of 25.4 mm, produced by the Beijing Glass Research Institute, was used. Silicone grease was used to couple the encapsulated LaBr$_3$(Ce) crystal with the R6233-100 PMT manufactured by Hamamatsu Photonics. The HPGe detector is a BE2020 planar germanium spectrometer manufactured by Canberra, with a thickness of 20 mm and a volume of 40000 mm$^3$. During the experiment, a $^{137}$Cs radioactive source generates 662 keV gamma rays through radioactive decay. The gamma rays undergo Compton scattering in the LaBr$_3$(Ce) crystal, generating Compton electrons, which are absorbed by the crystal, while some scattered photons escape the crystal and are absorbed by the nearby HPGe detector. The HPGe detector serves as a standard detector, used to indirectly determine the energy of the Compton electrons. By varying the angle $\theta$ between the $^{137}$Cs source, the LaBr$_3$(Ce) crystal, and the HPGe detector, coincidence events across a broad energy range are obtained. Figure~\ref{fig:2} depicts the schematic diagram of the data acquisition system used in the WACC experiment \cite{FPY}. The two signals from the crystal and HPGe detectors successively undergo low-threshold discrimination, delayed stretching, and a logical coincidence unit. The generated coincidence output signals are used as external trigger signals for the digitizer (model DT5751 from CAEN), which records the Compton coincidence events.



\begin{figure}[!htb]
\centering
\includegraphics
  [width=0.8\hsize]
  {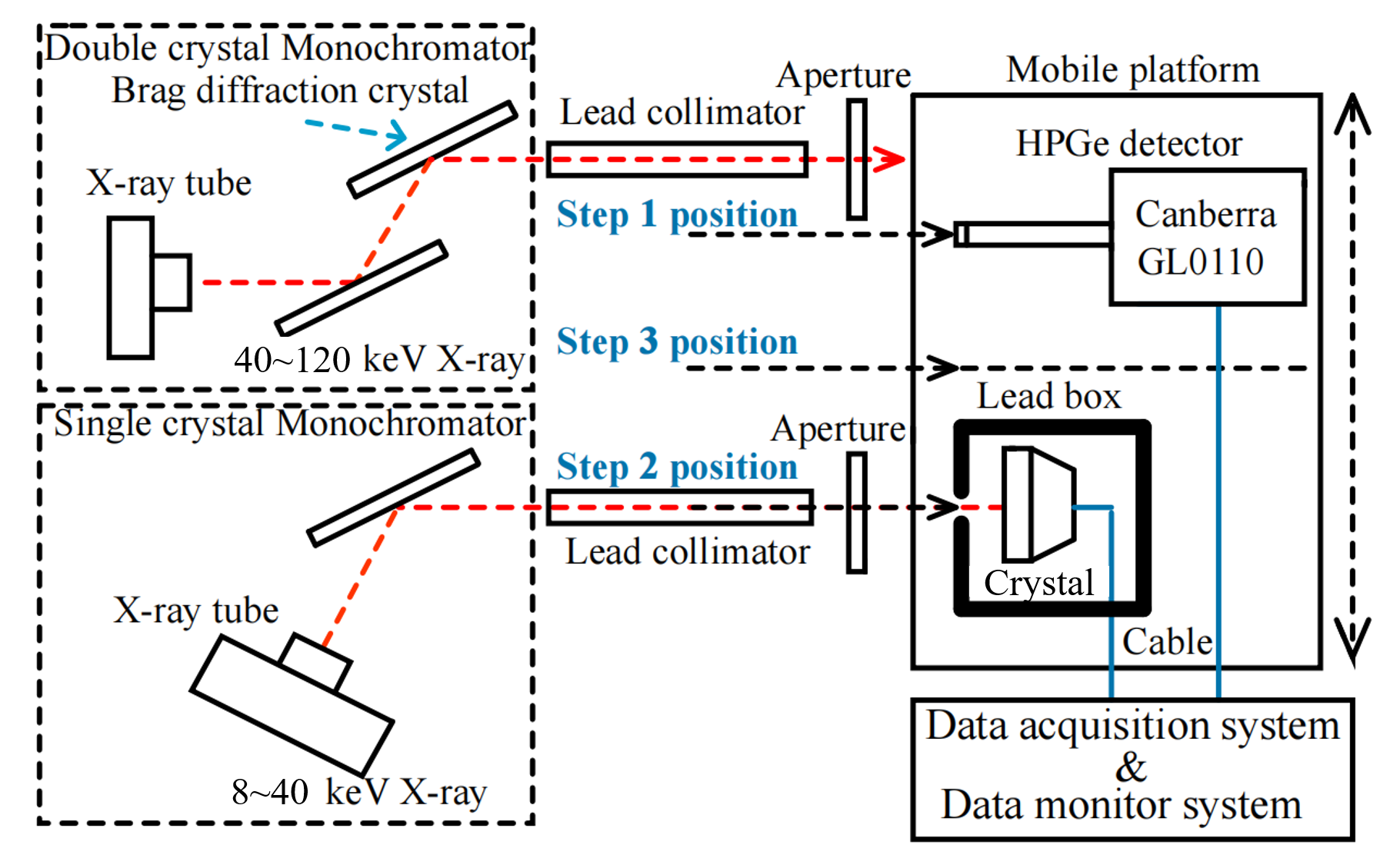}
\caption{Hard X-ray Calibration Facility \cite{feng2024detector}. Both the HPGe and crystal detectors are placed on a displacement platform and maintained on the same horizontal line.}
\label{fig:3}
\end{figure}

\begin{figure*}[!htb]
\centering
\includegraphics
  [width=\hsize]
  {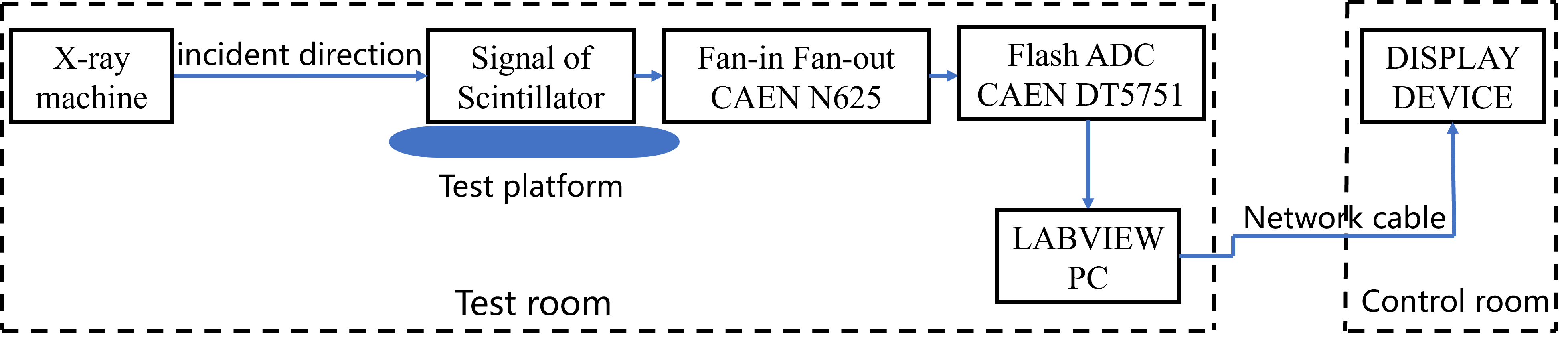}
\caption{Data acquisition system diagram for single-energy X-ray detection \cite{FPY}.}
\label{fig:4}
\end{figure*}

Figure~\ref{fig:3} depicts the HXCF experimental setup, which consists of four main components: an X-ray generator, a monochromator, a collimation system, and a standard detector \cite{FPY, 2021The, 2019The, zhang2022transition}. X-rays undergo Bragg diffraction with the crystal monochromator to produce monochromatic X-rays. The exit of the lead collimator is equipped with apertures of different sizes, and the lead shielding system can shield stray light coming out of the X-ray generator. The standard detector, a HPGe detector (Canberra GL0110), is used to determine the energy and flux of X-rays. After adjusting the X-ray beam current, the exit aperture is aligned with the HPGe detector, and data is collected for 300 seconds. The movable platform aligns the exit aperture with the LaBr$_3$(Ce) crystal, and data is also collected for 300 seconds. The LaBr$_3$(Ce) crystal is coupled with a PMT (Hamamatsu CR160 model) using Silicone grease. The data acquisition system, as shown in Fig.~\ref{fig:4}, uses the digitizer DT5751 to collect output signals \cite{FPY}. Finally, the X-ray exit aperture is moved to the middle position between the two detectors to obtain background data for 300 seconds.

Within the energy range we tested, the PMT operates in the proportional region and does not exhibit photon saturation effects. Hence, this study disregards the contribution of PMT non-proportionality to the energy resolution \cite{ZHANG2020162079}. The experimental introductions, testing procedures, and data collection for the WACC and HXCF experiments have been extensively described in the publication by our research team \cite{FPY}, so we only provide a brief overview in this paper. Upon completing these two experiments, we obtained the non-proportional response curves of the LaBr$_3$(Ce) crystal to Compton electrons and X-rays. In Section~\ref{chap:Evaluating the Contribution of Non-proportionality to Energy Resolution}, we describe how to evaluate the contribution of the non-proportionality to the total energy resolution by convolving the experimentally obtained non-proportional response curves with the secondary electron energies simulated by Geant4.

\subsection{PMT Single-Photoelectron Calibration}

PMTs exhibit high sensitivity and a favorable signal-to-noise ratio, providing significant advantages in detecting weak light signals. Accurate Single-Photoelectron Calibration (SPEC) of PMTs is crucial, as it directly impacts subsequent analyses of crystal energy resolution. This study employed Hamamatsu CR160 PMT and utilized LED-based calibration for single-photoelectron spectra \cite{wei2018consistency}.

\begin{figure}[!htb]
\centering
\includegraphics
  [width=0.8\hsize]
  {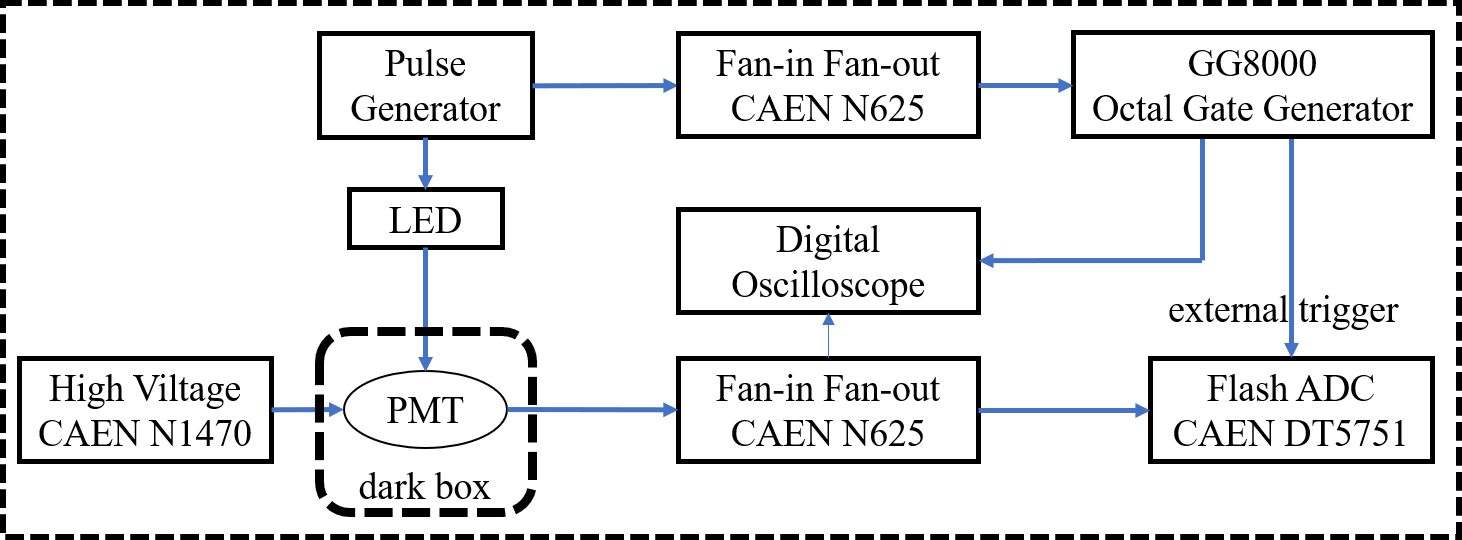}
\caption{Experimental setup schematic for single-photoelectron calibration.}
\label{fig:5}
\end{figure}

As shown in Fig.~\ref{fig:5}, the PMT is placed inside a dark box and powered by a high-voltage supply, model N1470 from CAEN. The pulse signal generator RIGOL DG1062Z drives the LED to emit faint blue-violet light and provides a positive synchronization signal. This positive synchronization signal first passes through the fan-in fan-out module (model N625 from CAEN), where it is converted to a negative polarity, and is then fed into the octal gate and delay generator (model GG8000 from Ortec) to adjust the gate width. When observing reasonable synchronization between this signal and the PMT signal on the oscilloscope, the synchronized signal is utilized as an external trigger for the data acquisition device DT5751 to eliminate the influence of dark noise. By altering the amplitude of the pulse signal, the LED's luminous intensity is adjusted to ensure that the incident light on the PMT's photocathode remains at the single-photon level.

Performing the aforementioned procedures, we obtained the single-photoelectron response of the Hamamatsu CR160 PMT at voltages ranging from –800 to –1800 V. Due to the insufficient gain of the Hamamatsu R6233-100 PMT used in the WACC experiment, we were unable to directly measure its single-photoelectron response and had to estimate it indirectly using the CR160 PMT. The basis of this indirect measurement method is the assumption that the crystal's absolute light output remains nearly constant. We used both PMTs to measure the $^{137}$Cs energy spectrum of the LaBr$_3$(Ce) crystal, and with the CR160 PMT already calibrated, we could estimate the single-photoelectron response of the R6233-100 PMT. According to the energy responses measured in the WACC and HXCF experiments, we calculated the absolute light yield of the LaBr$_3$(Ce) crystal and used it to evaluate the contributions of single-photoelectron fluctuations and statistical fluctuations to the total energy resolution.

\subsection{Experimental Setup for Radial Uniformity}

Different particle incidence positions lead to variations in photon generation locations within the crystal, resulting in discrepancies in the collection efficiency at the PMT's photocathode. The non-uniformity in photon generation within the crystal is categorized into radial ($X$- and $Y$-axes) and vertical ($Z$-axis) dimensions. The contribution of non-uniform light collection to the total energy resolution is assessed by performing Spot Scanning (SS) experiments and Geant4 optical simulations.

\begin{figure}[!htb]
\centering
\includegraphics
  [width=0.5\hsize]
  {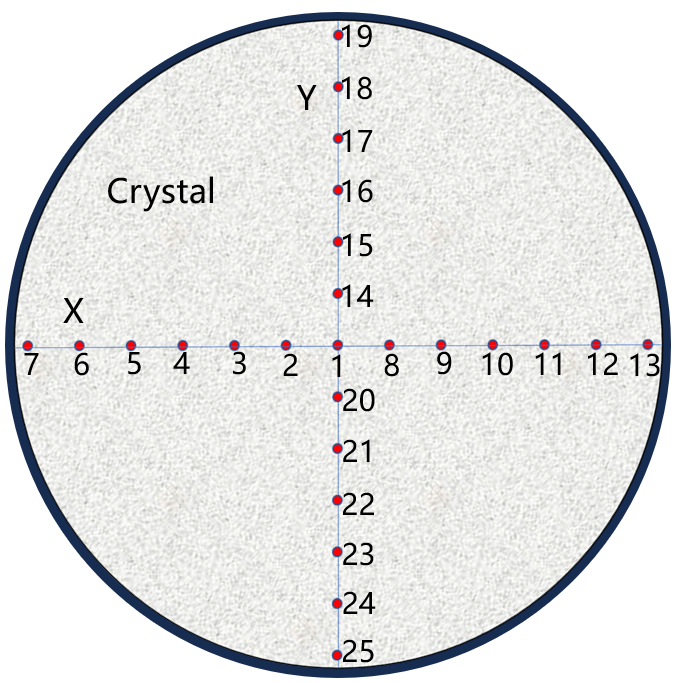}
\caption{Crystal's Be window entrance surface. Distribution and numbering of test points for assessing the positional non-uniformity of the crystals involve 25 designated locations.}
\label{fig:6}
\end{figure}

The HXCF (Fig.~\ref{fig:3}) was also utilized to test the position response of LaBr$_3$(Ce) crystal to 25 keV X-rays. Setting the coordinate parameters for each point, we adjusted the displacement platform to align the X-ray with the intended test positions. As depicted in Fig.~\ref{fig:6}, with the crystal center as the origin, test positions were distributed along the $X$- and $Y$-axes, with adjacent points spaced 2 mm apart. There were a total of 25 test positions, each with a statistical count of no less than 30,000. Background data were collected before and after the experiment for subtraction purpose.

\section{Monte Carlo simulation}

Geant4, developed by the European Organization for Nuclear Research (CERN) using C++ object-oriented technology, is a Monte Carlo application package employed for simulating the transport of particles within matter \cite{qian2021simulation, lu2022monte}.

\subsection{Simulation Approach}\label{chap:Simulation Approach}

Geant4 simulations allow for assessing the contributions of light yield non-proportionality and uneven light collection to the total energy resolution. Table~\ref{tab:G4 crystal} lists the crystal types and dimensions used in the simulations. Table~\ref{tab:G4 optical} presents the parameters utilized in the optical simulations. We completed three simulation tasks using Geant4, outlined as follows:


(1) We selected the $G4EmLivermorePhysics$ model to simulate the interactions of low-energy gamma photons with matter. Simulations of gamma emissions with energies of 25 keV, 60 keV, and 100 keV were performed, with 30,000 statistics for each energy point. These gamma photons interact with the LaBr$_3$(Ce) crystal, generating a series of secondary electrons. By tracking the trajectories of these secondary particles, we selected events with full energy deposition. The energies of the Compton electrons, photoelectrons, and Auger electrons in these full-energy deposition events were recorded. The energies of these secondary electrons were convolved with the electron response curve obtained from the WACC experiment to reconstruct the energy response distribution. The standard deviation of this distribution was used to evaluate the non-proportionality resolution of the crystal for gamma-ray detection.


(2) We selected the $G4EmLivermorePhysics$ model to simulate the interactions of low-energy electrons with matter. Simulations of electron emissions with energies of 25 keV, 60 keV, and 100 keV were performed, with 30,000 statistics for each energy point. We tracked the energy loss trajectories of the electrons in LaBr$_3$(Ce) crystal, recording the initial energy $E_{init}$ and deposited energy $E_{dep}$ for each step. The response for each step was calculated using the electron response curve obtained from the WACC experiment, which is the difference between the initial energy $E_{init}$ and the end energy ($E_{init}$ – $E_{dep}$). The responses of all steps were accumulated to obtain the response for a full-energy deposition event. After processing each event in this manner, the electron response distribution can be reconstructed, and the standard deviation of this distribution is used to evaluate the non-proportionality component of the crystal's electron resolution.


(3) We used the $G4OpticalSurface$ class to describe the boundary optical properties between materials, setting the boundary type to $dielectric\_dielectric$, the surface finish to $ground$, and the boundary calculation model to $unified$. In the $G4OpticalPhysics$ class, optical processes were set, including scintillation photon production, Cherenkov radiation, Burke absorption, Rayleigh scattering, and boundary processes (reflection, refraction, and absorption). We simulated a point source emitting 20,000 photons at 13 radial positions in LaBr$_3$(Ce) crystals at depths ($Z$-axis) of 1 mm, 2 mm, 3 mm, 4 mm, and 5 mm, covering a 4$\pi$ solid angle. The photon transport process was tracked, and the number of photons collected at the PMT photocathode was recorded to determine the collection efficiency non-uniformity of the LaBr$_3$(Ce) crystal at different positions.

\begin{table}[!htb]
\caption{Parameters of LaBr$_3$(Ce) crystal used in Geant4 simulations.}
\label{tab:G4 crystal}
\begin{tabular*}{15cm} {@{\extracolsep{\fill} } ccc}
\hline
Diameter & Density & Doping type \\
\hline
25.4 mm & 5.06 g/cm$^3$ & 5\% Ce$^{3+}$ \\
\hline
\end{tabular*}
\end{table}

\begin{table}[!htb]
\caption{Optical simulation parameters in Geant4 \cite{Epic, GFY}}.
\label{tab:G4 optical}
\begin{tabular*}{15cm} {@{\extracolsep{\fill} } cccc}
\hline
Material & Refractive index & Absorption length & Reflectivity \\
\hline
LaBr$_3$(Ce) & 1.9 & 100 cm & / \\
Quartz glass & 1.47 & 50 m & / \\
PMT glass & 1.54 & / & / \\
Teflon film & 1.35 & / & 95\% \\
\hline
\end{tabular*}
\end{table}

\subsection{Evaluating the Contribution of Non-proportionality to Energy Resolution}\label{chap:Evaluating the Contribution of Non-proportionality to Energy Resolution}

For full-energy deposition events, gamma rays interact with the LaBr$_3$(Ce) crystal to produce secondary electrons in two ways: (1) directly through a cascade process of photoelectric effect, generating multiple Auger electrons and characteristic X-rays, with the characteristic X-rays being reabsorbed by the crystal to generate secondary photoelectrons; (2) after multiple interactions (e.g., Compton scattering), one or more electrons are produced, and secondary photons undergo a cascade process of photoelectric effect.

For simulating the emission of 30,000 gamma photons (Section~\ref{chap:Simulation Approach}), full-energy deposition events are initially selected, followed by the selection of Compton electrons, photoelectrons, and Auger electrons from each full-energy deposition event. According to Equation~\ref{eq:Convolve}, the energy of the selected secondary electrons $E_e$ is convolved with the electron response curve $R_e(E_e)$ obtained from the WACC experiment to determine the energy response $L_\gamma$ of the incident gamma photons on the LaBr$_3$(Ce) crystal. In Equation~\ref{eq:Convolve}, $\Phi(E_\gamma,E_e)$ represents the distribution function of secondary electron energy. When a limited quantity of gamma photons are incident, Equation~\ref{eq:Convolve} can be rewritten as Equation~\ref{eq:L gamma}, employing discrete convolution to determine the energy response $L_{\gamma,i}$ of the i-th gamma photon. $M_i$ denotes the number of secondary electrons produced by the i-th gamma photon (i.e., the count of selected Compton electrons, photoelectrons, and Auger electrons). $E_{e,j}$ represents the energy of the j-th secondary electron. $R_e(E_{e,j})$ indicates the response of the j-th secondary electron, as provided by the electron response curve $R_e(E_e)$.

\begin{equation}\label{eq:Convolve}
L_\gamma = \int_0^\infty \Phi(E_\gamma,E_e)R_e(E_e)E_edE_e .
\end{equation}

\begin{equation}\label{eq:L gamma}
L_{\gamma,i} = \sum_{j=1}^{M_i}R_e(E_{e,j})E_{e,j} .
\end{equation}

According to Equation~\ref{eq:L gamma}, we calculate the energy response of each full-energy deposition event and reconstruct the energy response distribution of gamma photons. In Equation~\ref{eq:R gamma}, the calculated response $R_\gamma$ of gamma photons is expressed as the ratio of the centroid $\overline{L_\gamma}$ of the energy response distribution to the energy $E_\gamma$ of the incident gamma photons. Here, N represents the number of full-energy deposition events, and $L_{\gamma,i}$ is the energy response of the i-th full-energy deposition event. We calculate the standard deviation $\sigma$ of the energy response distribution according to Equation~\ref{eq:sigma gamma}. As shown in Equation~\ref{eq:delta gamma}, the ratio $\delta_{\gamma,non}$ of the standard deviation $\sigma$ to the centroid $\overline{L_\gamma}$ represents the non-proportionality contribution of LaBr$_3$(Ce) crystal to the total energy resolution for gamma rays.

\begin{equation}\label{eq:R gamma}
R_\gamma = \frac{\overline{L_\gamma}}{E_\gamma} = \frac{1}{E_\gamma} \frac{1}{N} \sum_{i=1}^NL_{\gamma,i} .
\end{equation}

\begin{equation}\label{eq:sigma gamma}
\sigma^2 = \frac{1}{N-1}\sum_{i=1}^N(L_{\gamma,i}-\overline{L_\gamma})^2 .
\end{equation}

\begin{equation}\label{eq:delta gamma}
\delta_{\gamma,non} = \frac{\sigma}{\overline{L_\gamma}} .
\end{equation}

When the incident particle is an electron, its energy loss process in LaBr$_3$(Ce) crystal is simpler than that of gamma photons, involving only scattering, ionization, and bremsstrahlung. For simulating the emission of 30,000 electron events (Section~\ref{chap:Simulation Approach}), we select the full-energy deposition events among them. According to Equation~\ref{eq:L electron}, we accumulate the energy response of all steps to obtain the energy response $L_{e,u}$ of the u-th full-energy deposition event. $M_u$ is the number of steps undergone by the u-th electron event to lose all energy. $E_{v,in}$ and $E_{v,out}$ are respectively the initial energy and end energy of the v-th step of the u-th electron event. $R_e$ is the electron response curve obtained from WACC experiments, where $R_e(E_{v,in})$ represents the response corresponding to the initial energy of the v-th step, and $R_e(E_{v,out})$ represents the response corresponding to the end energy of the v-th step.

\begin{equation}\label{eq:L electron}
L_{e,u} = \sum_{v=1}^{M_u}\{R_e(E_{v,in})E_{v,in} - R_e(E_{v,out})E_{v,out}\} .
\end{equation}

We calculate the energy response for each full-energy deposition event based on Equation~\ref{eq:L electron}, thereby reconstructing the energy response distribution of the electrons. In Equation~\ref{eq:R electron}, the calculated response $R_e$ of electrons is expressed as the ratio of the centroid $\overline{L_e}$ of the energy response distribution to the energy $E_e$ of the incident electrons. Here, N represents the number of full-energy deposition events, and $L_{e,u}$ is the energy response of the u-th full-energy deposition event. We calculate the standard deviation $\sigma$ of the energy response distribution according to Equation~\ref{eq:sigma electron}. As shown in Equation~\ref{eq:delta electron}, the ratio $\delta_{e,non}$ of the standard deviation $\sigma$ to the centroid $\overline{L_e}$ represents the non-proportionality contribution of LaBr$_3$(Ce) crystal to the total energy resolution for electrons. According to Equation~\ref{eq:resolution} and \ref{eq:intrinsic resolution}, we calculated the intrinsic resolution $\delta_{e,int}$ of the LaBr$_3$(Ce) crystal for electrons. We also analyzed the relationship between $\delta_{e,int}$ and $\delta_{e,non}$, discussing the impact of non-proportionality on the total energy resolution.

\begin{equation}\label{eq:R electron}
R_e = \frac{\overline{L_e}}{E_e} = \frac{1}{E_e} \frac{1}{N} \sum_{u=1}^NL_{e,u} .
\end{equation}

\begin{equation}\label{eq:sigma electron}
\sigma^2 = \frac{1}{N-1}\sum_{u=1}^N(L_{e,u}-\overline{L_e})^2 .
\end{equation}

\begin{equation}\label{eq:delta electron}
\delta_{e,non} = \frac{\sigma}{\overline{L_e}} .
\end{equation}

\section{Results}

This section presents the total energy resolution of the LaBr$_3$(Ce) crystal as well as the contribution of various components.

\subsection{Total Energy Resolution}

\begin{figure}[!htb]
\centering
\includegraphics
  [width=\hsize]
  {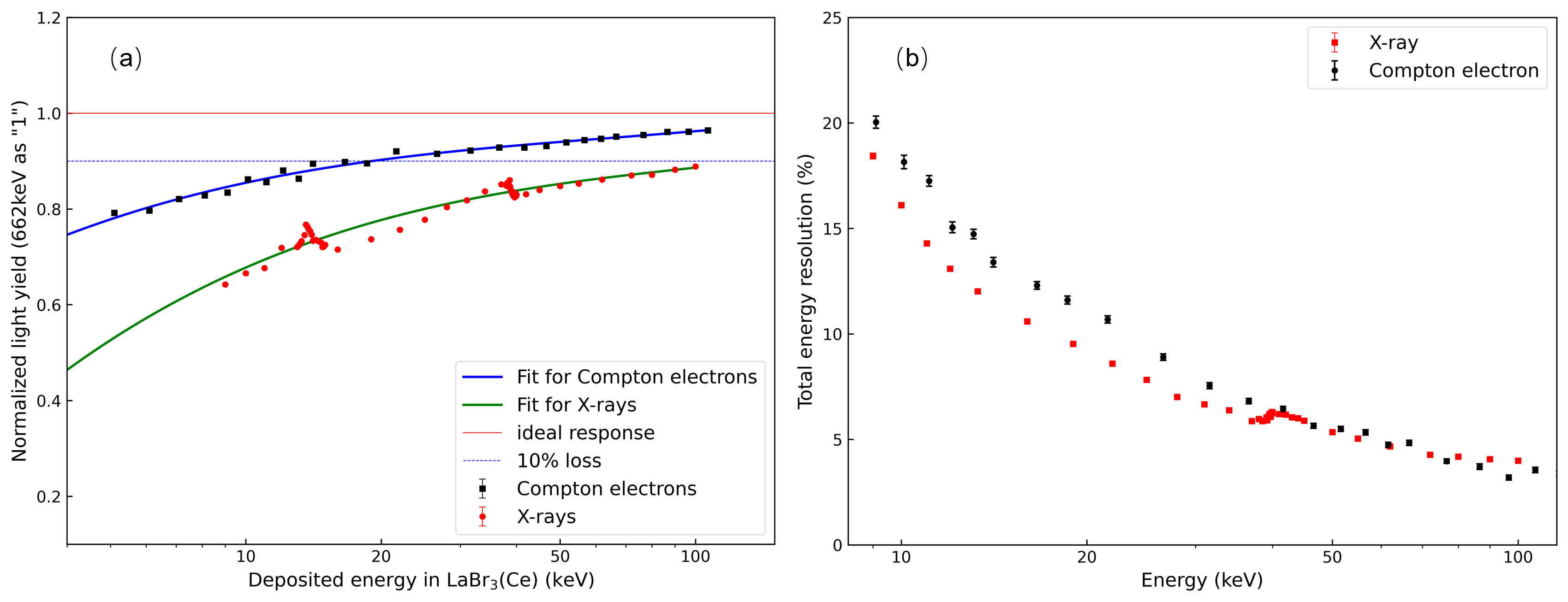}
\caption{(a) Energy response non-proportionality and fitting results of LaBr$_3$(Ce) crystal for X-rays and Compton electrons \cite{FPY}. (b) Total energy resolution of LaBr$_3$(Ce) crystal for X-rays and Compton electrons \cite{FPY}.}
\label{fig:7}
\end{figure}

In the article by our research team \cite{FPY}, the energy spectra of X-rays and Compton electrons for the LaBr$_3$(Ce) crystal have been measured using HXCF and WACC. We used previous experimental data to redraw the plots, and the results are shown in Fig.~\ref{fig:7} \cite{FPY}.
Figure~\ref{fig:7} (a) depicts the energy response non-proportionality of the LaBr$_3$(Ce) crystal for 9–100 keV X-rays and 5.1–106.6 keV Compton electrons. Figure~\ref{fig:7} (b) displays the energy resolutions of LaBr$_3$(Ce) crystal for 9–100 keV X-rays and 9.1–106.6 keV Compton electrons, represented by the standard deviation 1$\sigma$ of the full-energy peak. For 100 keV X-rays, the total energy resolution of LaBr$_3$(Ce) crystal is 3.99\% $\pm$ 0.04\%. Near the absorption edge (La's K-shell electron binding energy, 38.931 keV), the energy resolution deteriorates by no more than 0.5\%. For 100 keV Compton electrons, the total energy resolution of LaBr$_3$(Ce) crystal is 3.31\% $\pm$ 0.12\%. Near the La's K-shell electron binding energy, there is no deterioration in the electron's energy resolution.


\subsection{Single-photoelectron Fluctuations}\label{chap:Single-photoelectron Fluctuations}

\begin{figure}[!htb]
\centering
\includegraphics
  [width=\hsize]
  {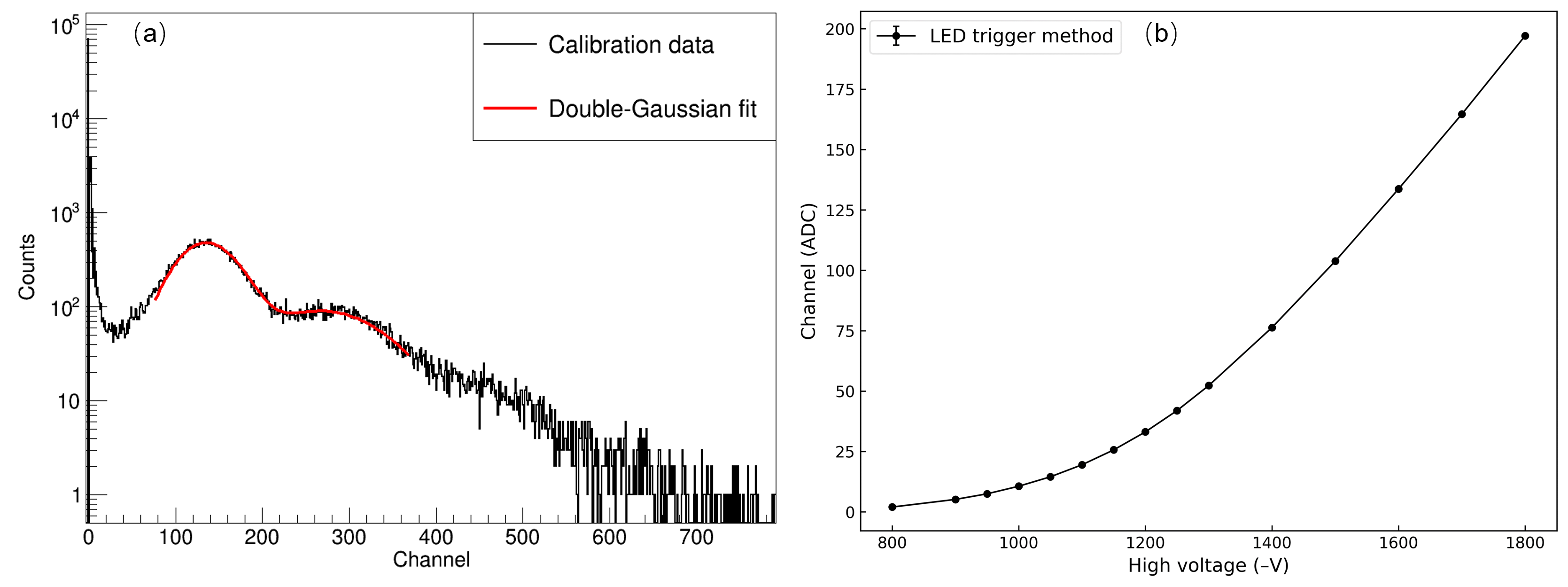}
\caption{(a) Photomultiplier Tube (PMT) CR160's single-photoelectron spectrum at –1600V operating voltage. (b) Single-photoelectron response of the Hamamatsu CR160 PMT at different operating voltages.}
\label{fig:8}
\end{figure}

\begin{figure}[!htb]
\centering
\includegraphics
  [width=\hsize]
  {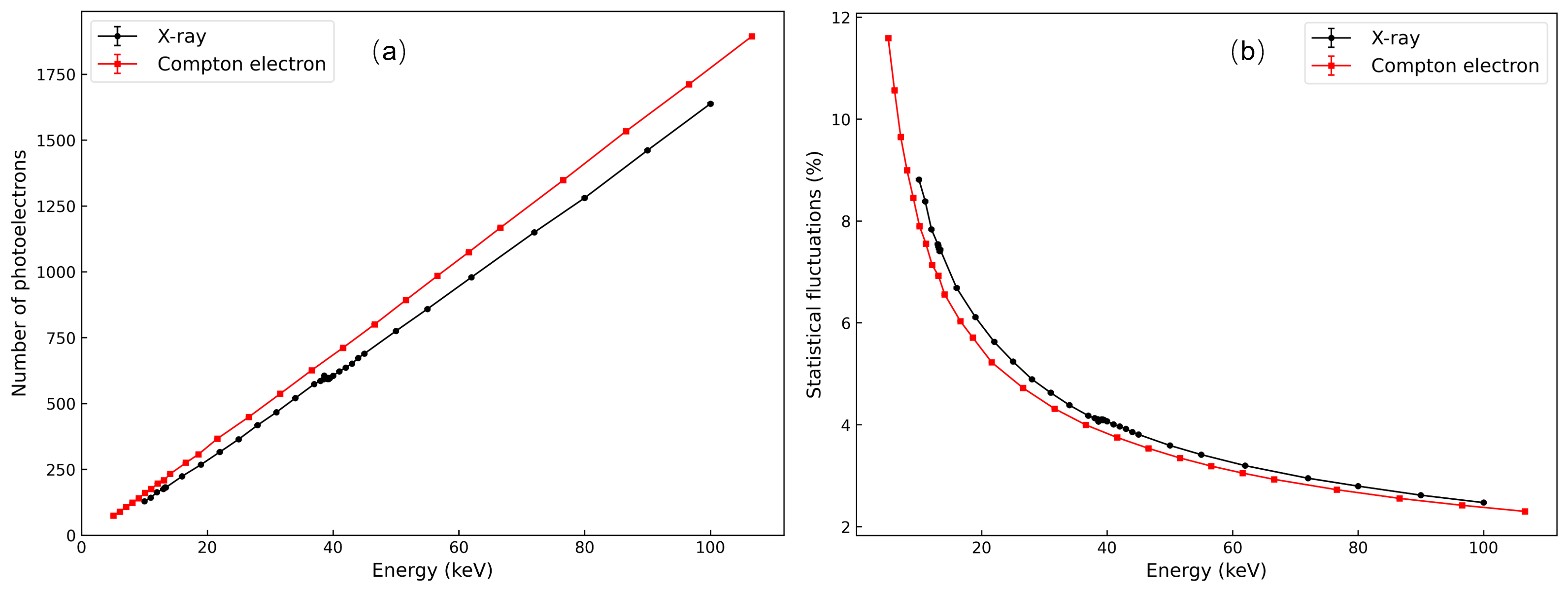}
\caption{(a) Absolute light yield of LaBr$_3$(Ce) crystal for 10–100 keV X-rays and 5.1–106.6 keV Compton electrons. (b) Statistical fluctuations of photoelectrons contribute to the total energy resolution when testing LaBr$_3$(Ce) with X-rays and Compton electrons.}
\label{fig:9}
\end{figure}

Figure~\ref{fig:8} (a) depicts the photoelectron spectrum of the CR160 PMT operating at –1600V, with the two peaks representing the single-photoelectron peak and the double-photoelectron peak respectively. A double-Gaussian fit was performed, and the fitting curve is shown in Fig.~\ref{fig:8} (a). By fitting the Gaussian, the centroid $\mu_i$ (i.e., the single-photoelectron response) and standard deviation $\sigma_i$ of the single-photoelectron peak can be obtained. The single-photoelectron resolution $(\sigma/E)_{spe}$ of the CR160 PMT is calculated by dividing $\sigma_i$ by $\mu_i$, yielding a value of 25.78\% ± 0.03\%. Figure~\ref{fig:8} (b) illustrates the results of single-photoelectron calibration, showing the single-photoelectron response at different voltages, revealing an exponential increase in the number of channels with voltage increment. The absolute light yield of the LaBr$_3$(Ce) crystal for 10–100 keV X-rays and 5.1–106.6 keV Compton electrons, represented by the symbol $N_{photoelectron}$, is obtained through single-photoelectron calibration, as shown in Fig.~\ref{fig:9} (a). According to Equation~\ref{eq:spe resolution}, for 100 keV X-rays, the contribution $\delta_{spe}$ of single-photoelectron fluctuations to the total energy resolution is calculated to be 0.64\% $\pm$ 0.07\%.

\begin{equation}\label{eq:spe resolution}
\delta_{spe} = \frac{(\sigma/E)_{spe}}{\sqrt{N_{photoelectron}}} .
\end{equation}


\subsection{Statistical Fluctuations}

The contribution of statistical fluctuations in the number of photoelectrons to the total energy resolution, denoted as $\delta_{st}$, is represented by Equation~\ref{eq:st resolution}. Here, $N_{photoelectron}$ stands for the number of photoelectrons obtained through single-photoelectron calibration of the PMT. 
Building upon our previous HXCF and WACC experiments, we calculated the $\delta_{st}$ for LaBr$_3$(Ce) crystal at each energy, as illustrated in Fig.~\ref{fig:9} (b). The contribution of statistical fluctuations in Compton electron testing was found to be lower than that of X-rays. We observed that statistical fluctuations in LaBr$_3$(Ce) crystal generally contribute significantly to the resolution. At 100 keV, they respectively reach 2.47\% $\pm$ 0.00\% for X-rays and 2.37\% $\pm$ 0.00\% for Compton electrons. During X-ray testing, there are jumps of no more than 0.1\% near the absorption edge.

\begin{equation}\label{eq:st resolution}
\delta_{st} = \frac{1}{\sqrt{N_{photoelectron}}}
\end{equation}


\subsection{Non-uniformity in Light Collection}

\begin{figure}[!htb]
\centering
\includegraphics
  [width=\hsize]
  {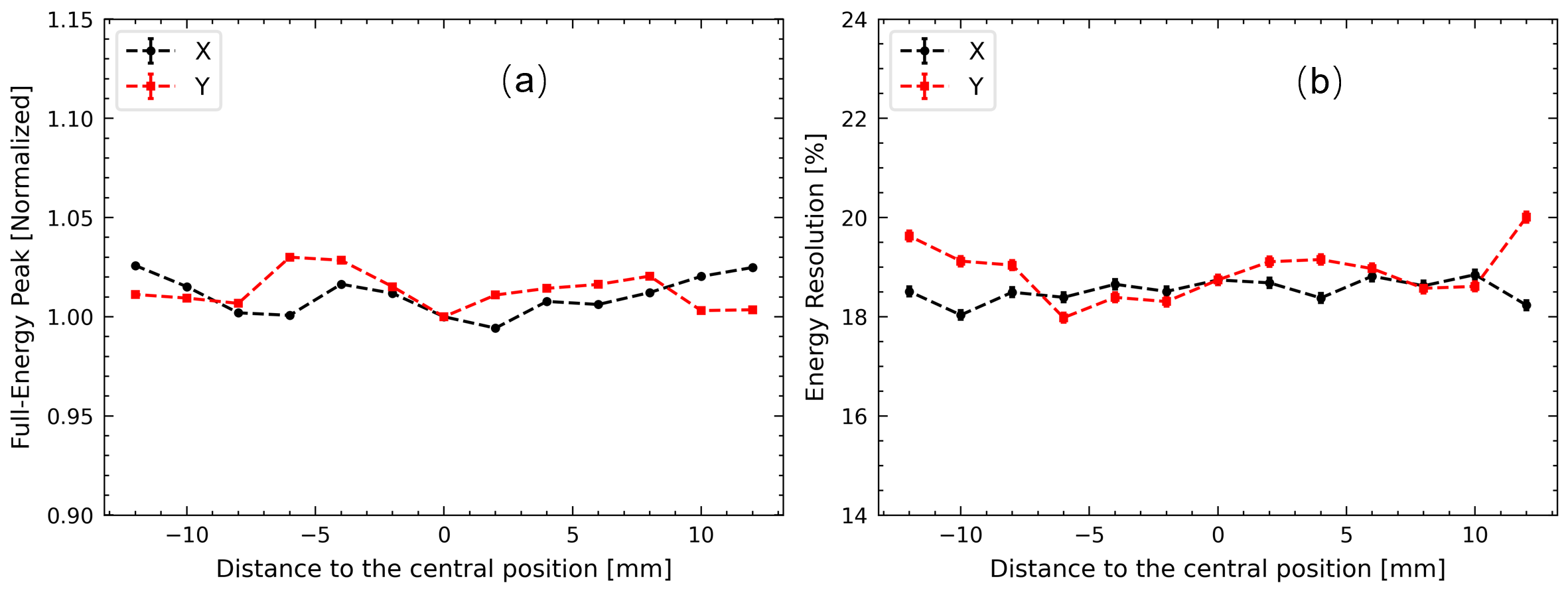}
\caption{(a) Relative full-energy peak of LaBr$_3$(Ce) crystal varies with position in the X and Y directions. (b) Energy resolution of LaBr$_3$(Ce) crystal varies with position in the X and Y directions.}
\label{fig:uneven}
\end{figure}

Due to varying light emission positions in the LaBr$_3$(Ce) crystal and the impact of uneven light collection, the measured full-energy peak position and energy resolution vary with the incident position of X-rays. Figures~\ref{fig:uneven} (a) and \ref{fig:uneven} (b) present the test results, showing fluctuations of 3.08\% for the peak position and 4.31\% for the energy resolution. The calculation method for these two values is the result of subtracting the minimum value from the maximum value and then dividing by the maximum value. To calculate the contribution of radial non-uniformity to the energy resolution, it is necessary to consider the weight of the measurement points' positions. We divide the $XY$ plane in Fig.~\ref{fig:6} into six concentric circles with radii of 2 mm, 4 mm, 6 mm, 8 mm, 10 mm, and 12 mm. We use the area of each region as the weight for the data points it contains. Based on Equation~\ref{eq:un resolution}, the contribution of radial non-uniformity to the total energy resolution, $\delta_{un}(radial)$, is calculated to be 0.18\% $\pm$ 0.01\%.


\begin{equation}\label{eq:un resolution}
\delta_{un} = \frac{\sigma}{mean} .
\end{equation}

\begin{figure}[!htb]
\centering
\includegraphics
  [width=0.6\hsize]
  {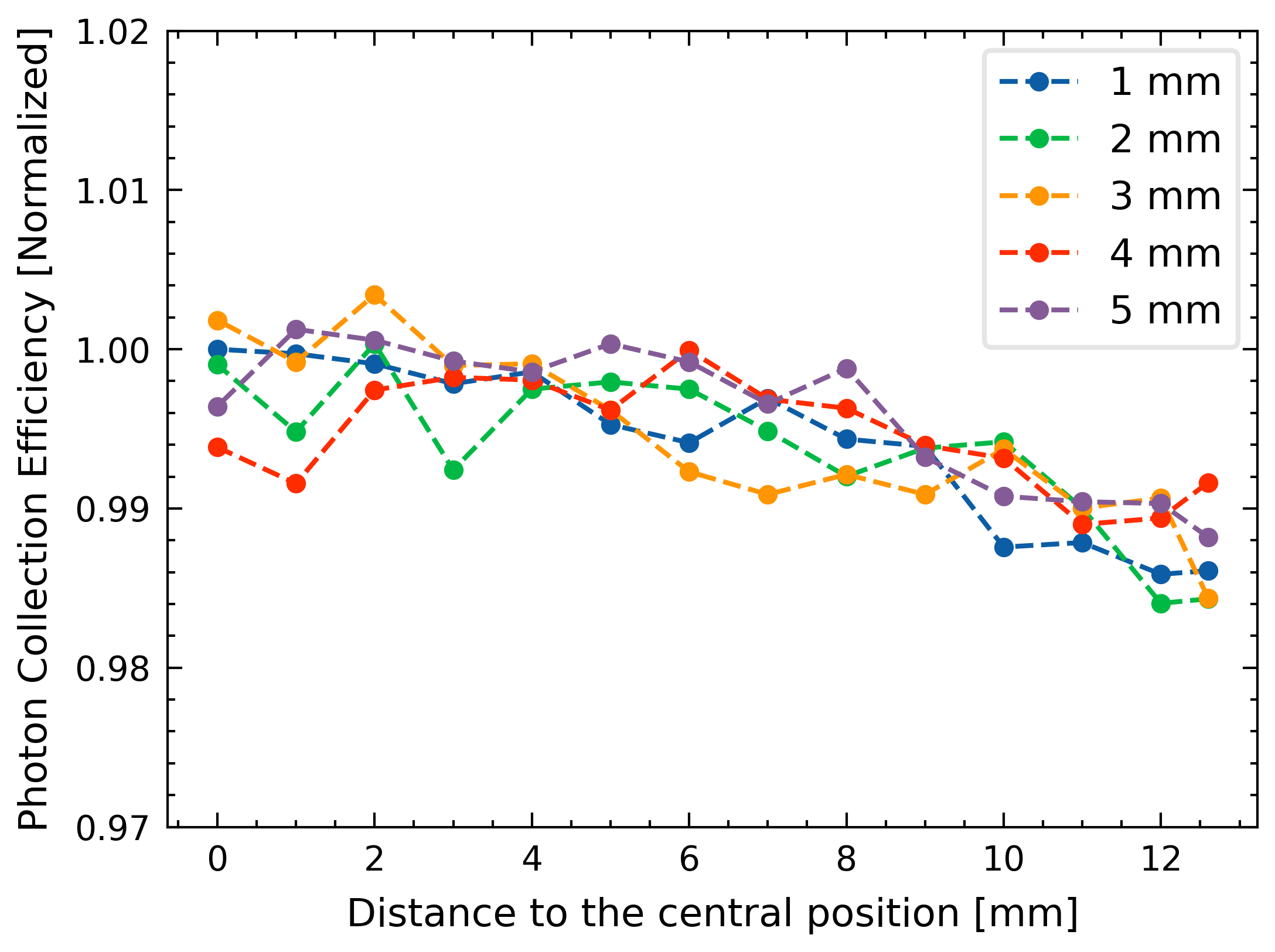}
\caption{Relative photon collection efficiency of LaBr$_3$(Ce) crystal varies with position.}
\label{fig:MC_Uneven_Eff}
\end{figure}

Geant4 simulations were conducted to analyze the variation in photon collection efficiency at different radial positions along the LaBr$_3$(Ce) crystal for depths ($Z$-axis) of 1 mm, 2 mm, 3 mm, 4 mm, and 5 mm. The results, depicted in Figure~\ref{fig:MC_Uneven_Eff}, are normalized to the 1 mm depth along the central axis. A decreasing trend in photon collection efficiency is observed from the LaBr$_3$(Ce) crystal's center towards its edges, showing an approximate decrease of 1.4\% at the periphery compared to the center. The simulation results exhibit a slight underestimate compared to the Spot Scanning experimental outcomes due to external factors influencing the testing. At a position emitting light at the central axis and a depth of 1 mm, the photon collection efficiency of the LaBr$_3$(Ce) crystal is approximately 88\%, with a relative variation of less than 1\% within the 5 mm depth range. To eliminate the influence of radial non-uniformity, the average of 14 data points for each depth was computed. After simplification, Equation~\ref{eq:un resolution} yields a contribution $\delta_{un}(vertical)$ of vertical non-uniformity for the LaBr$_3$(Ce) crystal of 0.08\% $\pm$ 0.00\%.

\subsection{The Contribution of Non-proportionality}

Both gamma photons and electrons ultimately result in secondary electron diffusion and energy deposition. To describe the electron response within a continuous energy range, empirical Equations~\ref{eq:NPR} and~\ref{eq:NPR x} were used to fit the Compton electron response obtained from the WACC experiment of the LaBr$_3$(Ce) crystal \cite{yang2022response}. Here, $E_e$ is the energy of electrons, $NPR(E_e)$ represents the electron response curve, and $P_n$ (n = 0,1,2,3) denotes fitting parameter. Merely fitting the energy range observed in the WACC experiment is insufficient, as Geant4 simulation results indicate that most secondary electrons generated by the cascade of photoelectric effects have energies mostly below a few keV. Therefore, extending the fitting curve to the low-energy range of 0–3 keV is necessary. The fitting results of the electron non-proportionality curve are presented in Fig.~\ref{fig:7} (a).

\begin{equation}\label{eq:NPR}
NPR(E_e) = P_0 + P_1 x + P_2 x^2 + P_3 x^3 .
\end{equation}

\begin{equation}\label{eq:NPR x}
x = log(E_e) .
\end{equation}

\begin{figure*}[!htb]
\centering
\includegraphics
  [width=\hsize]
  {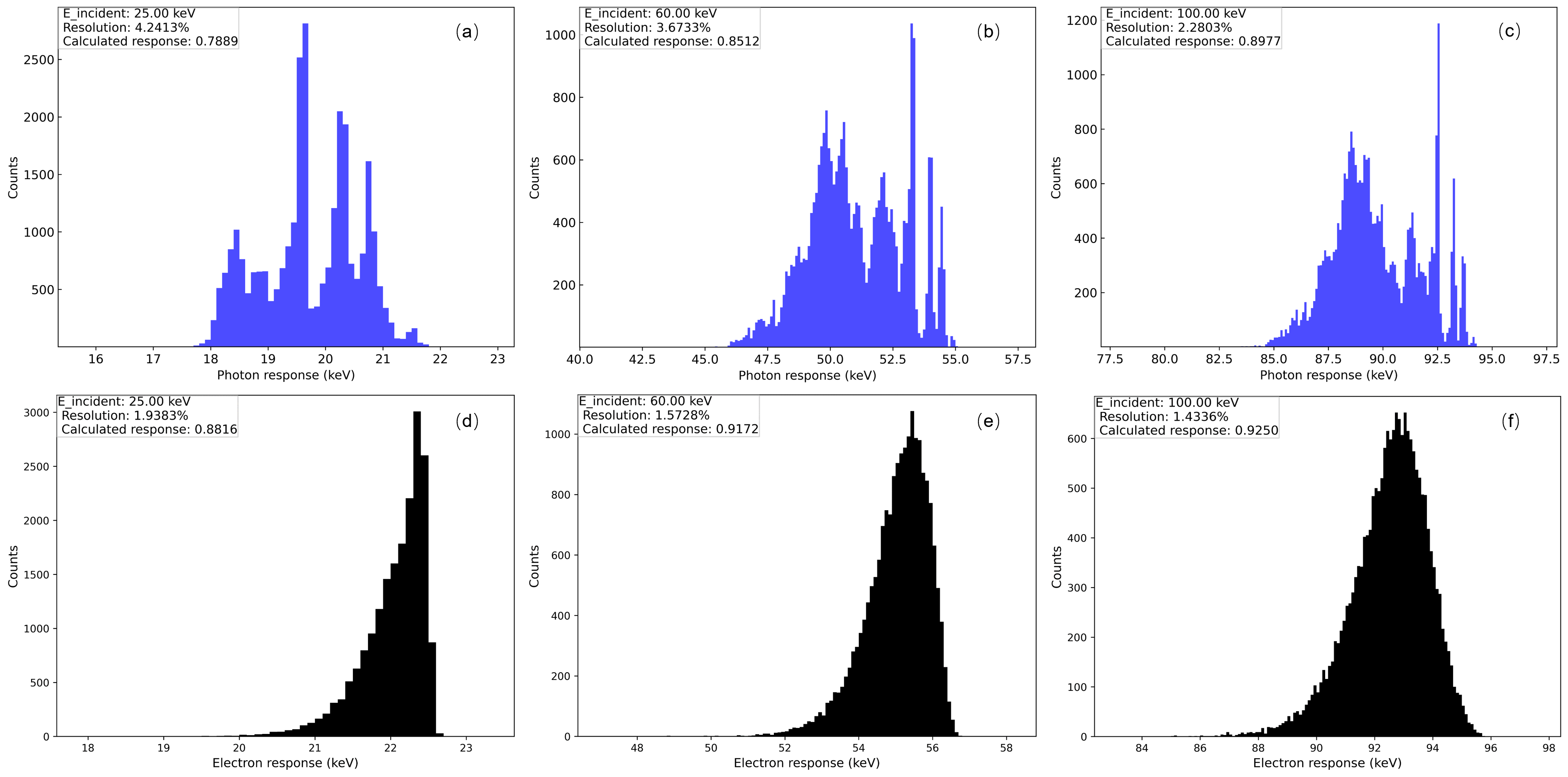}
\caption{The reconstructed energy response distributions for 25 keV (a), 60 keV (b), and 100 keV (c) gamma photons and 25 keV (d), 60 keV (e), and 100 keV (f) electrons are presented. The figures also display the statistical count of full-energy deposition events, the contribution of luminescent non-proportionality to the total energy resolution, and the reconstructed relative responses.}
\label{fig:12}
\end{figure*}

The fitted electron response curve was convolved with the secondary electron energies obtained from Geant4 simulations. Based on Section~\ref{chap:Evaluating the Contribution of Non-proportionality to Energy Resolution}, the reconstructed energy response distribution of gamma photons for different energies are depicted in Figures~\ref{fig:12} (a), (b), and (c), corresponding to 25 keV, 60 keV, and 100 keV, respectively. The annotations in the top-left corner of these figures indicate the incident particle energy (E\_incident), contribution of the non-proportionality component to the total energy resolution (Resolution), and calculated response (average response divided by incident energy). It's noteworthy that this distribution doesn't exhibit a Gaussian shape but rather displays some complex structures due to the diverse nature of secondary electron energies resulting from photon-matter interactions.
For 100 keV gamma photons, the calculated response $R_\gamma$ for the LaBr$_3$(Ce) crystal is 0.8977, falling short of 100 keV due to the deficient luminescence in the LaBr$_3$(Ce) crystal \cite{FPY}. At 100 keV, the contribution of luminescence non-proportionality to the total energy resolution is 2.28\% $\pm$ 0.00\%.


Based on the methodology outlined in Section~\ref{chap:Evaluating the Contribution of Non-proportionality to Energy Resolution}, the calculated responses for incident electrons at different energies are provided. Figures~\ref{fig:12} (d), (e), and (f) show the energy response distributions for incident energies of 25 keV, 60 keV, and 100 keV, respectively. For electrons at 100 keV, the calculated response $R_e$ of LaBr$_3$(Ce) crystal is 0.9250. Electron responses at the same energy surpass gamma photon responses due to differing interaction mechanisms between particles and matter, as elucidated in our previous energy response research \cite{FPY}. For 100 keV electrons, luminescence non-proportionality contributes 1.43\% $\pm$ 0.00\% to the total energy resolution.

\begin{figure}[!htb]
\centering
\includegraphics
  [width=0.6\hsize]
  {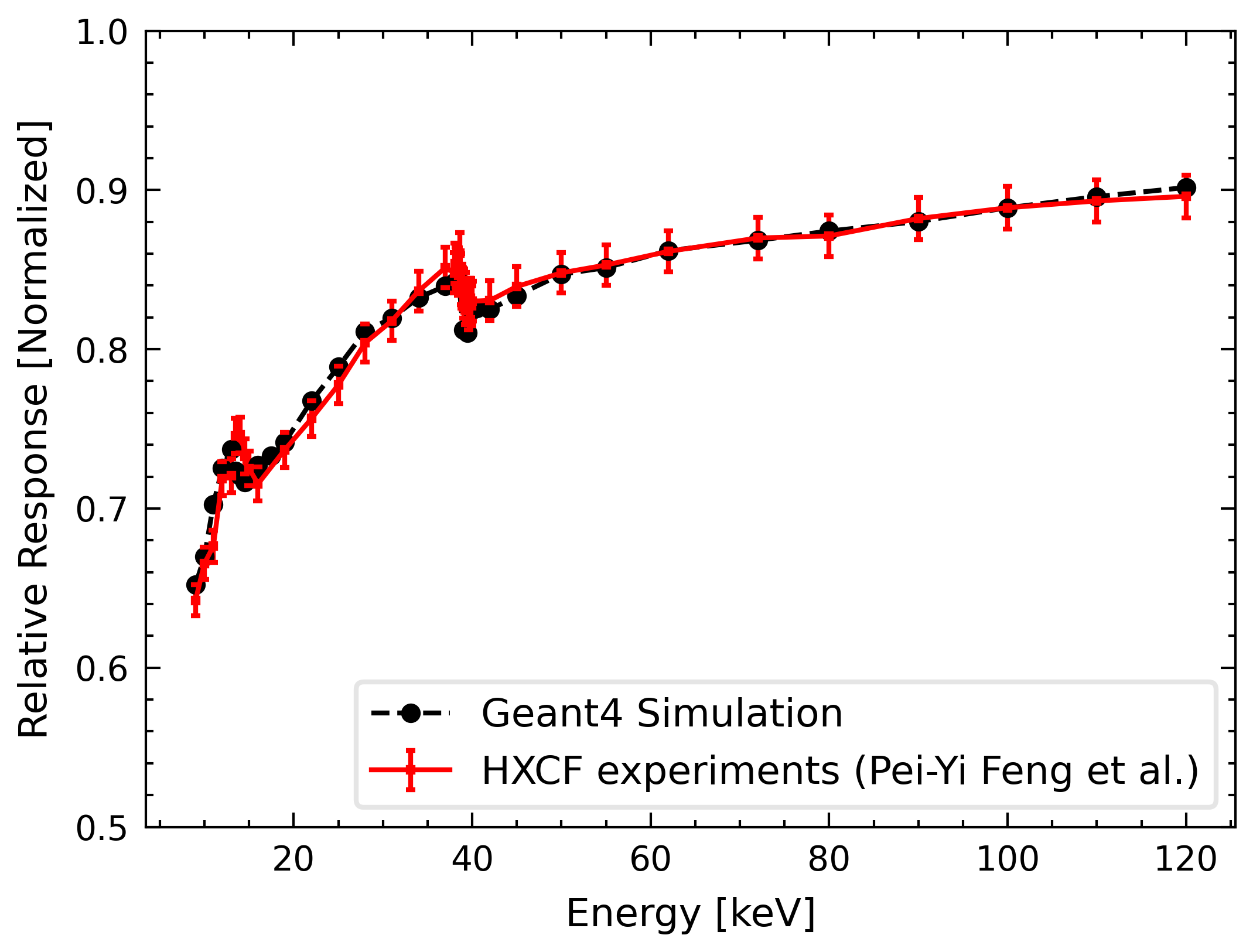}
\caption{Comparison between reconstructed photon response and experimentally obtained relative light yield.}
\label{fig:13}
\end{figure}

We successfully reconstructed the energy response distributions of gamma photons and electrons.  Due to the deficient luminescence of LaBr$_3$(Ce) crystal, the reconstructed responses do not exceed the energy of the incident particles, consistent with the findings of the HXCF and WACC experiments. We compared the calculated photon response with the reported relative light output of LaBr$_3$(Ce) crystal \cite{FPY}, as depicted in Fig.~\ref{fig:13}. The great consistency within the error range signifies the reliability of the Geant4 simulation results.


\subsection{Intrinsic Energy Resolution}

\begin{figure}[!htb]
\centering
\includegraphics
  [width=0.6\hsize]
  {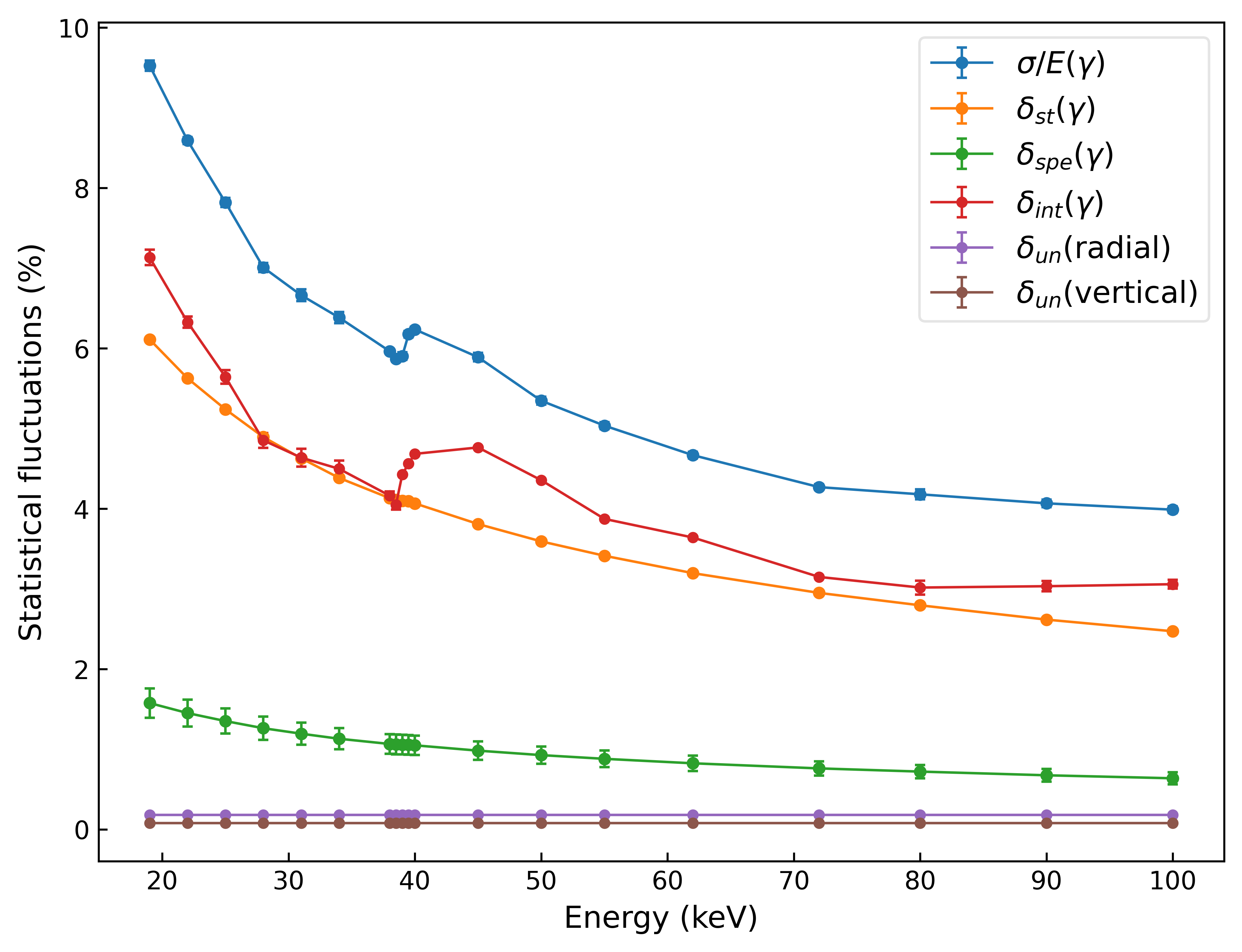}
\caption{Different contributions to the energy resolution of LaBr$_3$(Ce) crystal for 19-100 keV gamma rays.}
\label{fig:14}
\end{figure}

For 100 keV X-rays, the contribution $\delta_{trans}$ of energy transfer fluctuations is calculated as 2.04\% $\pm$ 0.08\% using Equation~\ref{eq:resolution} and the intrinsic resolution of the LaBr$_3$(Ce) crystal is determined as 3.06\% $\pm$ 0.06\% by Equation~\ref{eq:intrinsic resolution}. Table~\ref{tab:Components Resolution} provides a comprehensive breakdown of all factors contributing to the total energy resolution at 100 keV. Among the factors that may affect intrinsic resolution, the non-proportionality component has a significant impact. We consider luminescence non-proportionality to be an important source of intrinsic resolution. Additionally, energy transfer fluctuations, as an inherent property of the material, should not be overlooked. Figure~\ref{fig:14} shows the different contributions to the energy resolution of LaBr$_3$(Ce) crystal for 19-100 keV gamma rays. Statistical fluctuations and intrinsic resolution are the main components of the total energy resolution.


\begin{table}[!htb]
\caption{Contributions of individual components to the total energy resolution at 100 keV.}
\label{tab:Components Resolution}
\begin{threeparttable}
\begin{tabular*}{15cm} {@{\extracolsep{\fill} } ccc}
\hline
Component & Method & Result \\
\hline
$\sigma/E$ $(\gamma)$ & HXCF & 3.99 $\pm$ 0.04\% \\
$\sigma/E$ $(e)$ & WACC & 3.31 $\pm$ 0.12\% \\
$\delta_{spe}$ $(\gamma)$ & SPEC & 0.64 $\pm$ 0.07\% \\
$\delta_{spe}$ $(e)$ & SPEC & 0.61 $\pm$ 0.07\% \\
$\delta_{st}$ ($\gamma$) & SPEC + HXCF & 2.47 $\pm$ 0.00\% \\
$\delta_{st}$ (e) & SPEC + WACC & 2.37 $\pm$ 0.00\% \\
$\delta_{un}$ $(radial)$ & SS & 0.18 $\pm$ 0.01\% \\
$\delta_{un}$ $(vertical)$ & Geant4 & 0.08 $\pm$ 0.00\% \\
$\delta_{non}$ $(\gamma)$ & HXCF + Geant4 & 2.28 $\pm$ 0.00\% \\
$\delta_{non}$ $(e)$ & WACC + Geant4 & 1.43 $\pm$ 0.00\% \\
$\delta_{trans}$ & Equation~\ref{eq:resolution} & 2.04 $\pm$ 0.08\% \\
$\delta_{int}$ & Equation~\ref{eq:intrinsic resolution} & 3.06 $\pm$ 0.06\% \\

\hline
\end{tabular*}
\begin{tablenotes}
\item[*] Full forms of the abbreviations: Hard X-ray Calibration Facility (HXCF); Wide-Angle Compton Coincidence (WACC); Single-Photoelectron Calibration (SPEC); Spot Scanning experiments (SS).
\end{tablenotes}
\end{threeparttable}
\end{table}

\section{Conclusion}

Based on the published energy response of LaBr$_3$(Ce) crystal \cite{FPY}, we conducted a detailed study of its energy resolution. We analyzed six factors contributing to the energy resolution and quantified their contributions. We identified two main sources for the intrinsic resolution of LaBr$_3$(Ce) crystal: non-proportionality in luminescence and fluctuations in energy transfer to the luminescent center. For 100 keV X-rays, we calculated the non-proportionality component to be 2.28\% $\pm$ 0.00\% and the contribution of fluctuations in the energy transfer process to be 2.04\% $\pm$ 0.08\%. These two factors collectively result in an intrinsic resolution of 3.06\% $\pm$ 0.06\% for LaBr$_3$(Ce) crystal with 100 keV X-rays. Unlike the result reported by Deng Y. et al. for liquid scintillator (LAB + 2.5 g/L PPO + 3 mg/L bis-MSB) \cite{deng2022exploring}, the contribution of luminescence non-proportionality significantly affects the intrinsic resolution of LaBr$_3$(Ce) crystal. This indicates differences in the sources of intrinsic resolution among different types of scintillators. It is worth noting that deducting all considered external factors from the measured total energy resolution to acquire intrinsic resolution is a conservative approach, as the correlations between the components have not yet been studied. 
This study not only aids in comprehending the LaBr$_3$(Ce) detector performance but also represents a rare and precise measurement of the intrinsic resolution of LaBr$_3$(Ce) crystals internationally. Our research demonstrates significant originality in experimental design, data analysis, and measurement methodology, offering new insights into understanding intrinsic resolution.

\acknowledgments

This work was supported by the National Key Research and Development Program of China (Nos. 2022YFB3503600 and 2021YFA0718500), Strategic Priority Research Program of the Chinese Academy of Sciences (Nos. XDA15360102), and National Natural Science Foundation of China (Nos. 12273042 and 12075258).

\bibliographystyle{JHEP}
\bibliography{biblio}

\providecommand{\href}[2]{#2}\begingroup\raggedright\begin{thebibliography}{10}

\bibitem{canning2011first}
A.~Canning, A.~Chaudhry, R.~Boutchko et~al., \emph{First-principles study of luminescence in ce-doped inorganic scintillators}, \href{https://doi.org/10.1103/PhysRevB.83.125115}{\emph{Phys. Rev. B.} {\bfseries 83} (2011) 125115}.

\bibitem{kumar2009efficiency}
G.A.~Kumar, I.~Mazumdar and D.A.~Gothe, \emph{Efficiency calibration and simulation of a labr3 (ce) detector in close-geometry}, \href{https://doi.org/10.1016/j.nima.2009.08.045}{\emph{Nucl. Instrum. Methods Phys. Res. A.} {\bfseries 609} (2009) 183}.

\bibitem{mazumdar2013studying}
I.~Mazumdar, D.A.~Gothe, G.A.~Kumar et~al., \emph{Studying the properties and response of a large volume (946 cm3) labr3: Ce detector with $\gamma$-rays up to 22.5 mev}, \href{https://doi.org/10.1016/j.nima.2012.12.093}{\emph{Nucl. Instrum. Methods Phys. Res. A.} {\bfseries 705} (2013) 85}.

\bibitem{dhibar2018characterization}
M.~Dhibar, I.~Mazumdar, P.B.~Chavan et~al., \emph{Characterization of a 2$\times$ 2 array of large square bars of labr3: Ce detectors with $\gamma$-rays up to 22.5 mev}, \href{https://doi.org/10.1016/j.nima.2017.11.014}{\emph{Nucl. Instrum. Methods Phys. Res. A.} {\bfseries 883} (2018) 183}.

\bibitem{nassalski2006road}
A.~Nassalski, M.~Moszynski, T.~Szczesniak et~al., \emph{The road to the common pet/ct detector},  in \emph{2006 IEEE NSS/MIC}, vol.~3, pp.~1904--1908, IEEE, 2006, \href{https://doi.org/10.1109/NSSMIC.2006.354266}{DOI}.

\bibitem{gostojic2016characterization}
A.~Gostoji{\'c}, V.~Tatischeff, J.~Kiener et~al., \emph{Characterization of labr3: Ce and cebr3 calorimeter modules for 3d imaging in gamma-ray astronomy}, \href{https://doi.org/10.1016/j.nima.2016.06.044}{\emph{Nucl. Instrum. Methods Phys. Res. A.} {\bfseries 832} (2016) 24}.

\bibitem{dong2023development}
M.H.~Dong, Z.Y.~Yao and Y.S.~Xiao, \emph{Development and preliminary results of a large-pixel two-layer labr3 compton camera prototype}, \href{https://doi.org/10.1007/s41365-023-01273-5}{\emph{Nucl. Sci. Tech.} {\bfseries 34} (2023) 121}.

\bibitem{2023WW}
W.~Wang, X.L.~Li, J.H.~Wu et~al., \emph{Development and performance study of a dual-layer compton camera (in chinese)}, \href{https://doi.org/10.11889/j.0253-3219.2023.hjs.46.030401}{\emph{Nucl. Tech.} {\bfseries 46} (2023) 030401}.

\bibitem{kozyrev2016comparative}
A.~Kozyrev, I.~Mitrofanov, A.~Owens et~al., \emph{A comparative study of labr3 (ce3+) and cebr3 based gamma-ray spectrometers for planetary remote sensing applications}, \href{https://doi.org/10.1063/1.4958897}{\emph{Rev. Sci. Instrum.} {\bfseries 87} (2016) }.

\bibitem{wan2017research}
W.J.~Wan, W.~Zhou, J.B.~Zhou et~al., \emph{Research on neutron-gamma logging with an 241 am-be source based on labr 3: Ce detector}, \href{https://doi.org/10.11889/j.0253-3219.2017.hjs.40.110404}{\emph{Nucl. Tech.} {\bfseries 40} (2017) }.

\bibitem{chung1988environmental}
C.~Chung and C.J.~Lee, \emph{Environmental monitoring using a hpge-nai (tl) compton suppression spectrometer}, \href{https://doi.org/10.1016/0168-9002(88)90847-9}{\emph{Nucl. Instrum. Methods Phys. Res. A.} {\bfseries 273} (1988) 436}.

\bibitem{wen2021compact}
J.X.~Wen, X.T.~Zheng, J.D.~Yu et~al., \emph{Compact cubesat gamma-ray detector for grid mission}, \href{https://doi.org/10.1007/s41365-021-00937-4}{\emph{Nucl. Sci. Tech.} {\bfseries 32} (2021) 99}.

\bibitem{4545171}
M.~Moszyniski, A.~Nassalski, A.~Syntfeld-Kazuch et~al., \emph{Energy resolution of scintillation detectors—new observations}, \href{https://doi.org/10.1109/TNS.2007.908580}{\emph{IEEE Trans. Nucl. Sci.} {\bfseries 55} (2008) 1062}.

\bibitem{5076032}
L.~Swiderski, M.~Moszynski, A.~Nassalski et~al., \emph{Light yield non-proportionality and energy resolution of praseodymium doped luag scintillator}, \href{https://doi.org/10.1109/TNS.2009.2015590}{\emph{IEEE Trans. Nucl. Sci.} {\bfseries 56} (2009) 934}.

\bibitem{moszynski2002intrinsic}
M.~Moszynski, J.~Zalipska, M.~Balcerzyk et~al., \emph{Intrinsic energy resolution of nai (tl)}, \href{https://doi.org/10.1016/S0168-9002(01)01964-7}{\emph{Nucl. Instrum. Methods Phys. Res. A.} {\bfseries 484} (2002) 259}.

\bibitem{moszynski2004intrinsic}
M.~Moszynski, M.~Balcerzyk, W.~Czarnacki et~al., \emph{Intrinsic energy resolution and light yield nonproportionality of bgo}, \href{https://doi.org/10.1109/TNS.2004.829491}{\emph{IEEE Trans. Nucl. Sci.} {\bfseries 51} (2004) 1074}.

\bibitem{alekhin2013improvement}
M.S.~Alekhin, J.T.M.~De~Haas, I.V.~Khodyuk et~al., \emph{Improvement of $\gamma$-ray energy resolution of labr3: Ce3+ scintillation detectors by sr2+ and ca2+ co-doping}, \href{https://doi.org/10.1063/1.4803440}{\emph{Appl. Phys. Lett.} {\bfseries 102} (2013) }.

\bibitem{4545170}
W.W.~Moses, S.A.~Payne, W.S.~Choong et~al., \emph{Scintillator non-proportionality: Present understanding and future challenges}, \href{https://doi.org/10.1109/TNS.2008.922802}{\emph{IEEE Trans. Nucl. Sci.} {\bfseries 55} (2008) 1049}.

\bibitem{deng2022exploring}
Y.~Deng, X.L.~Sun, B.H.~Qi et~al., \emph{Exploring the intrinsic energy resolution of liquid scintillator to approximately 1 mev electrons}, \href{https://doi.org/10.1088/1748-0221/17/04/P04018}{\emph{JINST} {\bfseries 17} (2022) P04018}.

\bibitem{sriwongsa2017comparative}
K.~Sriwongsa, P.~Limkitjaroenporn and J.~Kaewkhao, \emph{Comparative study of light yield non-proportionality and energy resolution properties of ce-doped labr3 and luyap scintillator crystals}, \href{https://doi.org/10.1016/j.matpr.2017.06.165}{\emph{Materials Today: Proc.} {\bfseries 4} (2017) 6540}.

\bibitem{2017Intrinsic}
V.~Ranga, S.~Rawat, S.~Sharma et~al., \emph{Intrinsic resolution of compton electrons in cebr 3 scintillator using compact cct}, \href{https://doi.org/10.1109/TNS.2017.2779888}{\emph{IEEE Trans. Nucl. Sci.} {\bfseries 65} (2017) 616}.

\bibitem{kaintura2021energy}
S.S.~Kaintura, V.~Ranga, S.~Panwar et~al., \emph{Energy resolution of compton electrons in lacl 3: Ce using compact digitizer}, \href{https://doi.org/10.1007/s10967-021-07942-2}{\emph{J. Radioanal. Nucl. Chem.} {\bfseries 330} (2021) 1527}.

\bibitem{FPY}
P.Y.~Feng, X.L.~Sun, Z.H.~An et~al., \emph{The energy response of labr 3 (ce), labr 3 (ce, sr), and nai (tl) crystals for gecam}, \href{https://doi.org/10.12074/202312.00199}{\emph{Nucl. Sci. Tech.} {\bfseries 35} (2024) 23}.

\bibitem{2021The}
S.M.~Guo, J.J.~Wu and D.J.~Hou, \emph{The development,performances and applications of the monochromatic x-rays facilities in(0.218-301)kev at nim,china}, \href{https://doi.org/10.1007/s41365-021-00890-2}{\emph{Nucl. Sci. Tech.} {\bfseries 32} (2021) 14}.

\bibitem{2019The}
D.J.~Hou, J.J.~Wu, S.M.~Guo et~al., \emph{The realization and study of (21–301) kev monochromatic x-rays}, \href{https://doi.org/10.1016/j.nima.2019.02.024}{\emph{Nucl. Instrum. Methods Phys. Res. A.} {\bfseries 927} (2019) }.

\bibitem{zhang2022transition}
S.~Zhang, J.K.~Xia, T.~Sun et~al., \emph{Transition edge sensor-based detector: from x-ray to $\gamma$-ray}, \href{https://doi.org/10.1007/s41365-022-01071-5}{\emph{Nucl. Sci. Tech.} {\bfseries 33} (2022) 84}.

\bibitem{ZHANG2020162079}
H.Y.~Zhang, Y.H.~Yu, K.~Tariq et~al., \emph{Photomultiplier tube performance of the wcda++ in the lhaaso experiment}, \href{https://doi.org/10.1016/j.nima.2019.04.033}{\emph{Nucl. Instrum. Methods Phys. Res. A.} {\bfseries 958} (2020) 162079}.

\bibitem{feng2024detector}
P.Y.~Feng, Z.H.~An, D.L.~Zhang et~al., \emph{Detector performance of the gamma-ray transient monitor onboard dro-a satellite}, \href{https://doi.org/https://doi.org/10.1007/s11433-024-2458-9}{\emph{Sci. China-Phys. Mech. Astron.} {\bfseries 67} (2024) 1}.

\bibitem{wei2018consistency}
Y.T.~Wei, M.Y.~Guan, W.X.~Xiong et~al., \emph{Consistency test of pmt spe spectrum from dark-noise pulses and led low-intensity light}, \href{https://doi.org/10.1007/s41605-018-0042-6}{\emph{Rad. Detect. Tech. Methods} {\bfseries 2} (2018) 1}.

\bibitem{qian2021simulation}
X.L.~Qian, H.Y.~Sun, C.~Liu et~al., \emph{Simulation study on performance optimization of a prototype scintillation detector for the grandproto35 experiment}, \href{https://doi.org/10.1007/s41365-021-00882-2}{\emph{Nucl. Sci. Tech.} {\bfseries 32} (2021) 51}.

\bibitem{lu2022monte}
W.~Lu, L.~Wang, Y.~Yuan et~al., \emph{Monte carlo simulation for performance evaluation of detector model with a monolithic labr3 (ce) crystal and sipm array for $\gamma$ radiation imaging}, \href{https://doi.org/10.1007/s41365-022-01081-3}{\emph{Nucl. Sci. Tech.} {\bfseries 33} (2022) 107}.

\bibitem{Epic}
L.~Shanghai Shuojie Crystal Materials~Co., \emph{Labr3(ce) crystal (in chinese)}, {\emph{https://www.epic-crystal.com.cn/data/upload/20230817/64ddcb369e4dc.pdf} (2019) }.

\bibitem{GFY}
F.Y.~Guo, G.P.~Qu, X.L.~Sun et~al., \emph{The impact of scintillator geometry, surface treatment, and single- or double-sided readout on light collection in gamma-ray detectors (in chinese)}, \href{https://doi.org/https://doi.org/10.20173/j.cnki.ned.20240527.002}{\emph{Nucl. Electron. Detect. Tech.} (2024) }.

\bibitem{yang2022response}
G.C.~Yang, L.M.~Hua, F.~Lu et~al., \emph{Response functions of a 4 $\pi$ summing $\gamma$ detector in $\beta$-oslo method}, \href{https://doi.org/10.1007/s41365-022-01058-2}{\emph{Nucl. Sci. Tech.} {\bfseries 33} (2022) 68}.

\end{thebibliography}\endgroup

\end{document}